%Paper: gr-qc/9507059
%From: "Mohammad Khorrami" <MAMWAD@NETWARE2.IPM.AC.IR>
%Date: Mon, 31 Jul 1995 19:53:31 -305

\font\vlbf=cmbx10 scaled 1440
\font\lbf=cmbx10 scaled 1200
\baselineskip=1cm plus 0pt minus 0pt
\def\repageno{\footline={\hss\tenrm\folio\hss}}
\def\I{{\rm I}}
\def\II{{\rm II}}
\def\III{{\rm III}}
\nopagenumbers
\hfill{}

\vskip 3cm

\centerline{\vlbf A model universe with variable dimension:}

\centerline{\vlbf Expansion as decrumpling}

\hfill{}

\centerline{{\bf M. Khorrami}$^{1,2,3,*}$, {\bf R. Mansouri}$^{1,3,4}$,
   {\bf M. Mohazzab}$^{1,5}$, and {\bf M. R. Ejtehadi}$^{1,4}$}

\vskip 1cm
{\it
\noindent $^1$ Institute for Studies in Theoretical Physics and
           Mathematics, P.O.Box  5746, Tehran 19395, Iran.

\vskip -0.5cm
\noindent $^2$ Department of Physics, Tehran University,
             North-Kargar Ave. Tehran, Iran.

\vskip -0.5cm
\noindent $^3$ Institute for Advanced Studies in Basic Sciences,
             P.O.Box 159, Gava Zang, Zanjan 45195, Iran.

\vskip -0.5cm
\noindent $^4$ Department of Physics, Sharif University of Technology,
             P.O.Box 9161, Tehran 11365, Iran.

\vskip -0.5cm
\noindent $^5$ Department of Physics, Alzahra University,
             Tehran 19834, Iran.

\noindent $^*$ E-Mail: mamwad@irearn.bitnet}

\vskip 2cm

\centerline{\lbf Abstract}

\noindent We propose a model universe, in which the dimension of the space
is a continuous variable, which can take any real positive number. The
dynamics leads to a model in which the universe has no singularity. The
difference between our model and the standard Friedman-Robertson-Walker
models become effective for times much before the presently accepted age of
the universe.
\vfil\break
\repageno\pageno=1
\noindent{\lbf O Introduction}

\noindent The standard model of cosmology is based on the following
assumptions:

\item 1. Space-time is a differential manifold.
\item 2. Dimension of space-time is a fixed constant.
\item 3. Dimension of space-time is 3+1.
\item 4. Space-time is homogenous and isotropic.
\item 5. Expansion of the universe is adiabatic.
\item 6. Dynamics is based on the Einstein field equations.

Up to now there has been suggestions in the literature to modify somehow
the assumptions 3-6. Modification of the number of space-time to $D+1$, e.g.,
has been considered to account for inflation [1]. Inhomogenuos cosmological
models has been considered to study the growth of inhomogeneities in the
early universe or hoping to remove the big bang singularity [2]. In
homogenuos cosmology and in quantum cosmology the assumption of isotropy may
be abandoned [3,4]. But the most drastic change is the inflationary
paradigm, which means assuming nonadiabatic expansion of the universe [5,6].
This paradigm claims to remove almost all the deficiencies of the standard
model, except the singularity at the big bang. Lastly, some authors suggest
to modify the underlying dynamics of the general relativity with very
different motivations. Some like to remove the big bang singularity through
using quadratic lagrangians [7]; Brans-Dicke theory is another common
modifcation. Steady-State and Quasi-Steady-State theory through the
assumption of continuous creation of matter brings in another drastic change
in the common beliefs of the physicists [8]. It should be noted that none of
these modifications tackle the problem of quantum gravity.

We want to bring in a completely new picture for the space-time, and abandon
the first two assumptions of the standard model. Now, the observational
evidence for differentiability of space-time is actually very poor. In fact
the set of space-time events is not even continuous, and there are evidences
that the matter distribution in the universe, up to the present observed
limits of 100 h$^{-1}$ Mpc, is fractal or multifractal [9,10]. Such a
fractal sructure has been also observed in temperature fluctuations of the
cosmic background radiation observed by COBE [11]. Fractal structure of
space-time has also been used to interpret the quantum mechanics [12,13].
However, the dimension of space-time is always assumed to be a fix number,
usually 3+1. Authors using dynamical triangulation and Regge calculus in
general relativity or quantum gravity [14] don't change this assumption
either.

The picture we want to bring in cosmology is a generalization of polymeric
or tethered surfaces, which are in turn simple generalizations of linear
polymers to two-dimensionally connected networks [15,16]. Visualizing the
universe as a piece of paper, then the crumpled paper will stand for the
state of the early universe [17]. It should however be noted that the final
formulation of our model in this paper could as well be interpreted as a
generalization of fluid membranes [18]. In this case we can visualize the
universe as a clay; it can be like a three dimensional ball, or like a two
dimensional disc, or even like a one dimensional string. In each case the
effective dimension of the universe is a continuous number between the
dimension of the embedding space and some $D$ which could be 3. To study
the crumpling in the statistical physics one needs to define an embedding
space, which does not exist in our case. Therefore, we assume an embedding
space of arbitrary high dimension $\cal D$, which is allowed to be infinite.
This is neccesary, because the crumpling is highly dependent on the
dimension of embedding space.

To simplify our picture, we introduce a cosmological model with just the
space part having a continuosly varying dimension. We call this space, with
varying dimension, a D-space. Therefore we assume a homogeneuos and
isotropic universe, make a space-time decomposition, leaving  the time
coordinate unchanged. Now, we imagine our universe to be a D-space embedded
in a space with arbitrary large, maybe infinite, dimension $\cal D$. This
cosmic D-space consists of small cells with characteristic size of about the
Planck length, denoted by $\delta$. The cells, playing the role of the
monomers in polymerized surfaces, are allowed to have as many dimensions as
the embedding space. Therefore, the cosmic space can have a dimension as
large as the embedding space, like the polymers in crumpled phase. The
radius of gyration of the crumpled cosmic space should play the role of the
Friedman parameter of a FRW cosmology in $D+1$ dimensional space-time, where
$D$ is the fractal dimension of the crumpled space in the embedding space
and could be as high as $\cal D$. The expansion of space is understood now
as decrumpling of cosmic space. In the course of decrumpling the fractal
dimension of space changes from $\cal D$ to $D$, where $D$ is about three.
To formulate the problem we write down the Hilbert-Einstein action for a FRW
metric in $D$ dimension. Now the Friedman parameter $a$, and the dimension
$D$ are both dynamical variables. The dynamical property of $D$ could lead
to difficulties if the model were not homogenuos [19], and we had to
consider a Lagrangian density in the action. The above mentioned cell
structure of the universe brings in the next simplification which is a
relation between $a$ and $D$. Our model system becomes again a system with
one degree of freedom, but the field equations are more complex. It turns
out that these generalized field equation admits the FRW model as a limit.

\hfill{}

\noindent{\lbf I The action}

\noindent We begin with a $D+1$ dimensional space-time $M\times R$, where
$M$ is assumed to be homogeneous and isotropic. The space-time metric is
written as
$$ds^2=-dt^2+a^2(t)\delta_{ij}dx^idx^j.\eqno{(\I .1)}$$
For simplicity, throughout this paper we take $M$ to be flat. The Ricci
scalar is then calculated to be
$$R=2D{d\over{dt}}\Big({{\dot a}\over a}\Big)+D(D+1)\Big({{\dot a}\over a}
   \Big)^2.\eqno{(\I .2)}$$
It is then easily seen that
$$R\sqrt{\vert g\vert}=-D(D-1)\Big({{\dot a}\over a}\Big)^2+
  \hbox{total time derivative}.\eqno{(\I .3)}$$
Note that we are still considering $D$ to be constant. The homogeneity of
the metric allows one to integrate the gravitational Lagrangian density.
Therefore, the gravitational Lagrangian bocomes
$$L^{(0)}_G=-{1\over{2\kappa}}D(D-1)\Big({{\dot a}\over a}\Big)^2a^D,
  \eqno{(\I .4)}$$
where $a^D$ is the volume of $M$.

Coupling this Lagrangian to the source
$$L^{(0)}_M={1\over 2}\theta^{\mu\nu}g_{\mu\nu}, \eqno{(\I .5)}$$
gives the complete Lagrangian of the gravitational field. In (I.5),
$$T^{\mu\nu}:={1\over\sqrt{\vert g\vert}}\theta^{\mu\nu},\eqno{(\I. 6)}$$
and
$$\eqalign{T^{00}&=\rho ,\cr T^{0i}&=0,\cr T^{ij}&=g^{ij}p=a^{-2}p
  \delta_{ij},\cr}\eqno{(\I .7)}$$
where $\rho$ and $p$ are the energy density and pressure, respectively.
(I.7) then leads to
$$\theta^{00}=\tilde\rho :=\rho a^D,\eqno{(\I .8)}$$
and
$$\theta^{ij}=\tilde p\delta^{ij}:=p a^{D-2}\delta^{ij}.\eqno{(\I .9)}$$
Note that, as in the $3+1$ dimensional case, to get the correct field
equations, in varying the action with respect to $a$, we have to take
$\tilde\rho$ and $\tilde p$ as constants [20]. The complete Lagrangian is
then
$$L^{(0)}=-{1\over{2\kappa}}D(D-1)\Big({{\dot a}\over a}\Big)^2 a^D+\Big(
  -{{\tilde\rho}\over 2}+{{\tilde p D a^2}\over 2}\Big) .\eqno{(\I .10)}$$

This Lagrangian suffers from the fact that its dimension is not constant.
Note that, we have assumed that the dimension of $\kappa$ is constant, that
is (Length)$^{D_0-1}$. To make a Lagrangian with a constant dimension, we
multiply the above Lagrangian by $a_0^{D_0-D}$, where $a_0$ is a quantity
with the dimension of length, in fact the scale of the universe when the
dimension is $D_0$. Clearly, this factor brings no change in General
Relativity, where $D=D_0=3$. Now, for our general case of variability of the
space dimension, the constant part of this factor, $a_0^{D_0}$, can be
ommited. Therefore, we arrive at the Lagrangian
$${L\over{a_0^{D_0}}}=-{{D(D-1)}\over{2\kappa}}\Big({{\dot a}\over a}\Big)^2
   \Big({a\over{a_0}}\Big)^D+\Big( -{{\hat\rho}\over 2}+{{\hat p D a^2}\over
   2}\Big)=:{\cal L},\eqno{(\I .11)}$$
where,
$$\eqalign{\hat\rho&:=\rho\Big({a\over{a_0}}\Big)^D,\cr
    \hat p&:=p a^{-2}\Big({a\over{a_0}}\Big)^D,\cr}\eqno{(\I .12)}$$
We take this Lagrangian as our starting point. To check this Lagrangian,
we derive the corresponding field equation for $D=D_0={\rm const.}$; one
expects to obtain the familiar Friedman equations [21], which are
$$\eqalign{\Big({{\dot a}\over a}\Big)^2&={{2\kappa\rho}\over{D_0(D_0-1)}},
   \cr {{\ddot a}\over a}&=-{\kappa\over{D_0(D_0-1)}}\big[ (D_0-1)\rho -D_0p)
   \big] .\cr}\eqno{(\I .13)}$$
Now, the Lagrangian (I.11) leads to the following field equation
$$-(D_0-1)\bigg[{{\ddot a}\over a}+{{D_0-2}\over 2}\Big({{\dot a}\over a}
  \Big)^2\bigg] =\kappa p,\eqno{(\I .14)}$$
which is just a combination of the two equations (I.13). The other equation
comes from the continuity relation
$${d\over{dt}}(\rho a^{D_0})+p{d\over{dt}}(a^{D_0})=0.\eqno{(\I .15)}$$
It is then seen that the Lagrangian (I.11), together with the continuity
relation (I.15) completes the usual picture of Friedman-Robertson-Walker
cosmology.

To implement the idea of variability of space dimension, we assume a
cellular structure for space: the universe consists of $N$ $D_0$-dimensional
cells. In a fictive embedding space, as far as $N$ is finite, there is no
$D$-dimensional arrangement of the cells with non-zero volume. So we assume
the cells to have an arbitrary number of extra dimensions, each of which has
a length scale $\delta$. Then the following relation holds between the $D$-
and $D_0$-dimensional valume of the cells.
$${\rm vol}_D({\rm cell})={\rm vol}_{D_0}({\rm cell})\delta^{D-D_0}.
   \eqno{(\I .16)}$$
Our $D$-dimensional universe will have an effective lenght scale,
corresponding to the radius of gyration of a crumpled surface [22], equal to
$a$. Hence we may write
$$\eqalign{a^D&=N{\rm vol}_D({\rm cell})\cr &=N{\rm vol}_{D_0}({\rm cell})
   \delta^{D-D_0}\cr &=a_0^{D_0}\delta^{D-D_0},\cr}\eqno{(\I .17)}$$
or
$$\Big({a\over\delta}\Big)^D=e^C,\eqno{(\I .18)}$$
where $C$ is a constant. This is an important constraint, which relates
the length scale of the universe to its dimension: as $D$ grows up to
infinity, $a$ decreases down to $\delta$ (but not less than it), and as $a$
grows, $D$ decreases. In other words, the expansion of the universe is
through reduction of its dimension.

Using this constraint, the equation of motion becomes
$$(D-1)\bigg\{ {{\ddot a}\over a}+\Big[ {{D^2}\over{2D_0}}-1-
  {{D(2D-1)}\over{2C(D-1)}}\Big]\Big({{\dot a}\over a}\Big)^2\bigg\} +
  \kappa p\Big( 1-{D\over{2C}}\Big) =0.\eqno{(\I .19)}$$
Using (I.18), one gets
$$a{{dD}\over{da}}=-{{D^2}\over C}.\eqno{(\I .20)}$$
Once again, we can check whether the equations (I.19) and (I.20) are
consistent with the Einstein field equation (I.14). From (I.20), it is
seen that in the limit $C\to\infty$, $D=D_0={\rm const.}$
In this case, the equation (I.19) becomes the same as (I.14). Hence, the
standard Friedman cosmology is the $C\to\infty$ limit of our model.
Moreover, assuming that $C\gg 1$, it is seen that in the vicinity of $D_0$,
our model universe behaves like a Friedman model. Had we generalized the
Lagrangian (I.10) by multiplying it by the other possibility
($\delta^{D_0-D}$), we would have gotten another field equation, which would
not lead to the familiar Friedman equation. Therefore, the generalization
(I.11) is unique.

The field equation (I.19) is not sufficient to obtain $a$. A continuity
equation, and an equation of state, are also needed. A dimensional reasoning
leads to the following generalization of the continuity equation (I.15):
$${d\over{dt}}(\rho a^D a_0^{D_0-D})+p{d\over{dt}}(a^D a_0^{D_0-D})=0,
   \eqno{(\I .21)}$$
or
$${d\over{dt}}\bigg[\rho\Big({a\over{a_0}}\Big)^D\bigg] +p{d\over{dt}}
      \Big({a\over{a_0}}\Big)^D=0.\eqno{(\I .22)}$$
Adding an equation of state to (I.18), (I.19), and (I.22), the formulation of
the dynamics of our model universe is complete.

\hfill{}

\noindent{\lbf II Qualitative behavior of the system}

\noindent Our system is defined through (I.18), (I.19), (I.22), and an
equation of state. This system is more difficult to be solved analytically.
To understand its qualitative behavior, we find a first integral of motion
and study its properties.

{}From the Lagrangian (I.11), one can define a Hamiltonian:
$$\eqalign{{\cal H}:&=\dot a{{\partial\cal L}\over{\partial\dot a}}-{\cal L},
  \cr &=\dot u{{\partial\cal L}\over{\partial\dot u}}-{\cal L},\cr
  &=-{{D(D-1)}\over{2\kappa}}e^C\Big({\delta\over a_0}\Big)^D\dot u^2-
  \Big({{\hat p}\over 2}D\delta^2e^{2u}-{{\hat\rho}\over 2}\Big) .\cr}
  \eqno{(\II .1)}$$
where $u$ is defined through
$$u:=\ln{a\over\delta}={C\over D}.\eqno{(\II .2)}$$
Using
$${{d\cal H}\over{dt}}=-{{\partial\cal L}\over{\partial t}},
   \eqno{(\II .3)}$$
we have
$${d\over{dt}}\bigg[ -{{D(D-1)}\over{2\kappa}}e^C
  \Big({\delta\over a_0}\Big)^D\dot u^2-{{\hat p}\over 2}D\delta^2 e^{2u}
  \bigg] =-{D\over 2}\delta^2 e^{2u}{{d\hat p}\over{dt}},\eqno{(\II .4)}$$
or
$${d\over{du}}\bigg[ {{D(D-1)}\over{2\kappa}}\Big({\delta\over a_0}\Big)^D
  \dot u^2\bigg] +p\Big({\delta\over a_0}\Big)^D\Big( D-{{D^2}\over{2C}}\Big)
  =0.\eqno{(\II .5)}$$
Note that $\hat\rho$ and $\hat p$ are considered as sources. The equation
(II.5) is equivalent to
$${{C^2}\over{2\kappa}}{d\over{dD}}\Big({{D-1}\over{D^3}}
   e^{-{{CD}\over D_0}}\dot D^2\Big) +pe^{-{{CD}\over D_0}}\Big({1\over 2}-
   {C\over D}\Big) =0.\eqno{(\II .6)}$$
This means that the function
$${\cal U}(D):=\int^D dD'\; p(D')e^{-{{CD'}\over D_0}}\Big({1\over 2}-
  {C\over D'}\Big) ,\eqno{(\II .7)}$$
serves as a potential for the kinetic energy
$${\cal T}:={{C^2}\over{2\kappa}}\Big({{D-1}\over{D^3}}
  e^{-{{CD}\over D_0}}\dot D^2\Big) .\eqno{(\II .8)}$$

Now, we will show that the system described by these, has two turning
points. To do so, we must consider the behavior of the pressure. For large
values of $D$, it is natural to use a radiation equation of state:
$$p={1\over D}\rho .\eqno{(\II .9)}$$
{}From this, and the equation of continuity (I.22), one can calculate the
pressure as
$$p=p_0 e^{C( {D\over D_0}-1)}
  \Big({D\over D_0}\Big)^{{C\over D_0}-1}.\eqno{(\II .10)}$$
It is then easily seen that at large $D$'s,
$${\cal U}(D)\sim {{D_0}\over{2C}}e^{-C}\Big({D\over D_0}\Big)^{C\over D_0}.
  \eqno{(\II .11)}$$
The point $D=2C$ is the point where the potential attains its minimum. For
$D$ near zero, assuming that the pressure remains finite (nonzero), it is
seen that
$${\cal U}(D)\sim -C\ln D,\eqno{(\II .12)}$$
that is, $\cal U$ grows unboundedly to infinity at $D\to\infty$, as well as
$D\to 0$. This means that there are two turning points, one above $D=2C$,
the other below it.

However, the kinetic term (II.8) changes sign at $D=1$. The above discussion
is valid, provided ${\cal T}\ge 0$. So, to have two turning points, the
constant ${\cal E}:={\cal U}+{\cal T}$ must be sufficiently low to make the
lower turning point greater than 1. By the way, dimension of the universe
less than 1 means that it is a disconnected set of cells, which we are not
going to consider it.

\hfill{}

\noindent{\lbf III Behavior of the model near the lower turning point of the
dimension}

Taking $D_0$ as the lower turning point of the potential, we have, from
(II.6),
$${{C^2}\over{2\kappa}}{{D-1}\over{D^3}}e^{-{{CD}\over{D_0}}}\dot D^2 +
  \int_{D_0}^D dD'\; p(D')e^{-{{CD'}\over{D_0}}}\Big({1\over 2}-{C\over D'}
  \Big) =0.\eqno{(\III .1)}$$
Now, defining
$$\epsilon :=D-D_0,\eqno{(\III .2)}$$
and using (II.10), we have, to lowest orders in $\epsilon$ and
$\dot\epsilon$,
$$\dot\epsilon^2 =4 A\epsilon ,\eqno{(\III .3)}$$
where
$$\bar A:={{\kappa D_0^2 p_0}\over{2C(D_0-1)}}.\eqno{(\III .4)}$$
This approximation holds wherever
$${{C\epsilon}\over{D_0}}\ll 1.\eqno{(\III .5)}$$
It is now easy to solve this equation to obtain
$$\epsilon =A(\tau -t)^2.\eqno{(\III .6)}$$
We use this expression for $t\le\tau$, where $\tau$ is the
lower-turning-point time. Here, all the times are measured from the big bang
point of the standard cosmology. Now, we want to show that, even down to
Planck's time the variations in $D$ are negligible, and one can essentially
take $D=D_0$. We have
$$\eqalign{ {{\epsilon (t_i)}\over{\epsilon (t_f)}}=:{{\epsilon_i}\over
  {\epsilon_f}}&=\Big({{\tau -t_i}\over{\tau -t_f}}\Big)
   ^2\cr &=(1+A^{1/2}T\epsilon_f^{-1/2})^2\cr &=:(1+\alpha )^2,\cr}
   \eqno{(\III .7)}$$
where
$$T:=t_f-t_i,\eqno{(\III .8)}$$
$t_f$ is the present time, and $t_i$ is some initial time, when we want to
calculate the value of $\epsilon$. To estimate the second term of the
parentheses, we use equations of standard cosmology, during the radiation
era:
$$\eqalign{p&={1\over 3}\rho\cr &={1\over 3}\rho_P
   \Big({{t_P}\over{t_f}}\Big)^2.\cr}\eqno{(\III .9)}$$
So we have
$$\eqalign{\alpha&=\Big({{9\kappa p_0}\over{4C}}\Big)^{1/2}T
   \epsilon_f^{-1/2},\cr&=\Big({{3\kappa\rho_P}\over{4C}}\Big)^{1/2}
   {t_P\over{t_f}}T\epsilon_f^{-1/2}.\cr}\eqno{(\III .10)}$$
Now using $\delta\sim t_P$, $a_0\sim t_f$, $t_f\sim 10^{17} {\rm s}$
(the age of the universe), $\rho_P\sim 10^{93} \rm kg\; m^{-3}$ [21],
$m_P^4\sim 10^{97} \rm kg\; m^{-3}$, and $D_0=3$, we obtain
$$\eqalign{C&={1\over{D_0}}\ln{{t_f}\over{t_P}}\cr &\sim 600,\cr}
  \eqno{(\III .11)}$$
and
$$\alpha\sim 10^{-3}{T\over{t_f}}\epsilon_f^{-1/2}.\eqno{(\III .12)}$$
If $\epsilon_f$ is small (less than $10^{-6}$), $\alpha$ will be greater
than one. So
$$\eqalign{\epsilon_i&\sim\epsilon_f\alpha^2\cr &\sim 10^{-6}
  \Big({T\over{t_f}}\Big)^2.\cr}\eqno{(\III .13)}$$
Comparing this with the criterion (III .5), we see that our approximation
holds, and $\epsilon_i$ is small, for
$$T\ll 100t,\eqno{(\III .14)}$$
that is, the dimension of the cosmos has been constant from at least
10 times the "standard age of universe" before the "Big Bang".

\hfill{}

\noindent{\lbf IV Conclusions}

\noindent The model we are proposing is qualitatively different from
hietherto considered theories with extra dimensions, such as Kaluza-Klein
theories [23], supergravity theories [23], and superstring theories [24].
There the 'external' and 'internal' dimensions are fixed, and the internal
space being compactified, is of the size of Planck length. Therefore any
change in dynamics comes from the change in the Lagrangian and not because
of the variability of the dimension. We, in contrast, take the dimension as
a dynamical variable. The picture we are using could be that of a
decrumpling 3-dimensional space-membrane. This picture has led us to a model
universe with a dynamics wich depends on the dimension of D-space. As we go
back in time more and more, the dependence on the dimension becomes more
effective. However, there is no beginning of time, and no Big Bang.
Therefore, we consider the time of Big Bang in the standard model as a
relative zero point of time! The higher turning point, where the dimension
of D-space is more than 1000, could be considered as a beginning of the
decrumpling or the expansion of the universe, but we should be aware that
this point should not be considered as the 'creation'-time in the sence of
standard model. In fact, our model does not have any real starting time,
because it is an oscillating model. As our model doesn't have any starting
time, the traditional horizon problem in standard cosmology does not show up
in our model.

The most exciting feature of our model seems to be the absence of any
singularity. Even, contrary to our initial expectations, the dimension of
universe remains finite.

It should be noted that despite resolving the problems of the standard
model, we could still have inflation within the decrumpling universe model.
The impact of our model on the structure formation, nucleosynthesis,
flatness problem, and dark matter remains to be considered. We are currently
studying the other problems of the standard model and will turn to them in
future publications.

\vfil\break

\noindent{\lbf References}

\item {[1]} M. Szydlowski, M. Biesiada, Phys. Rev. {\bf D41} (1988) 2487.
\item {[2]} E. Ruiz, J. M. M. Senovilla, Phys. Rev. {\bf D45} (1992) 1995.
\item {[3]} J. J. Halliwel, Proceedings of the Jerusalem winter school on
  Quantum Cosmology and Baby Universe (1990).
\item {[4]} M. P. Ryan, L. C. Shepley, Homogeneous Relativistic Cosmologies,
  Princeton Uviversity Press (Princeton, N. J. 1975).
\item {[5]} A. H. Guth. Phys. Rev. {\bf D23} (1981) 347.
\item {[6]} A. Linde, Lectures on Inflationary Cosmology, hep-th/9410082.
\item {[7]} V. Mukhanov, R. H. Brandenberger, Phys. Rev. Lett. {\bf 68}
  (1992) 1969.
\item {[8]} F. Hoyle, V. Narlikar, Rev. Mod. Phys.,   (1995).
\item {[9]} P. H. Coleman, L. Pietronero, Phys. Reps. {\bf 213} (1992) 311.
\item {[10]} X. Luo, D. N. Schramm, Science {\bf 256} (1992) 513.
\item {[11]} E. M. de Gauveia Dal Pino et. al., Evidence for a very
  large-scale fractal structure in the universe from COBE measurements,
  preprint, 1995.
\item {[12]} S. Amir-Azizi, A. J. G. Hey, T. R. Morris, Quantum Fractals
  Complex Systems {\bf 1} (1987) 923.
\item {[13]} L. Nottale, Chaos, Solitons \& Fractals, {\bf 4} (1994) 361.
\item {[14]} A. A. Migdal, J. G. P., {\bf 5} (1988) 711.
\item {[15]} D. Nelson, T. Piran, and S. Weinberg, Statistical Mechanics of
  Membranes and Surfaces, World Scientific ( Singapore, 1989).
\item {[16]} F. F. Abraham, M. Kardar, Science {\bf 252} (1991) 419.
\item {[17]} See ref. 15, p. 115.
\item {[18]} See ref. 15, p. 46.
\item {[19]} D. Hochberg, J. T. Wheeler, Phys. Rev. {\bf D43} (1991) 2617.
\item {[20]}
\item {[21]} E. W. Kolb, M. S. Turner, The early universe, Addison-Wesley
  Publishing Company (1990).
\item {[22]} P. J. de Gennes, Scaling concepts in polymer physics, Cornell
  University Press, (Ithaca and London, 1973).
\item {[23]} M. J. Duff, B. E. Nilsson, and C. N. Pope, Phys. Rep.
  {\bf 130,1} (1986).
\item {[24]} J. H. Schwarz, Superstrings, World Scientific (1975).
\end